\documentclass[useAMS,usenatbib,twocolumn]{mn2e}
\usepackage{graphicx}
\usepackage{dcolumn}   
\usepackage{bm}        
\usepackage{amssymb}   

\def\bq{\begin{equation}}
\def\eq{\end{equation}}
\def\bqy{\begin{eqnarray}}
\def\eqy{\end{eqnarray}}

\def\ga{\gamma}

\def\la{\lambda}

\def\cale{\mathcal{E}}

\def\calo{\mathcal{O}}

\begin{document}

\title[Veltmann models with a cosmological constant]{The effects of a non-zero cosmological constant on the Veltmann models}
\author[M. Lingam]{Manasvi Lingam$^{1}$\thanks{E-mail:
manasvi@physics.utexas.edu}\\
$^{1}$Institute for Fusion Studies, The University of Texas, Austin, TX 78712, USA\\}

\date{}

\pagerange{\pageref{firstpage}--\pageref{lastpage}} \pubyear{2014}

\maketitle

\label{firstpage}

\date{\today}

\begin{abstract}
The Veltmann models, which include the Plummer and Hernquist models as special cases, are studied in the presence of a cosmological constant. Physically relevant quantities such as the velocity dispersion profiles and the anisotropy parameter are computed through the use of the self-consistent approach. The cutoff radii for these models and the mass contained within this volume are also calculated. It is shown that the inclusion of a cosmological constant leads to many observable quantities such as the surface density, dispersion profiles and the anisotropy parameter becoming increasingly modified. In some scenarios, they are easily distinguished from the case where the cosmological constant is absent, as a result of their non-monotonic behaviour. The effects of neighbouring gravitational systems on the central system are also studied, and compared against the effects arising from the cosmological constant. Consequently, it is suggested that the effects of a cosmological constant can prove to be quite important when modelling dilute collisionless systems.

\end{abstract}


\section{Introduction}
The self-consistent approach is widely used in astrophysics because it can be used to compute physically relevant quantities such as the velocity dispersion profiles and the virial theorem. However, one of the underlying difficulties is that it is very difficult to obtain exact solutions, even when the simplest possible scenario of spherical symmetry is assumed. The most commonly used approach is to pick a potential--density pair, and then work backwards to determine the distribution function \citep{pl11,ed16, ly62, os79, ja83,b200,bi87, he90}. It is also possible to adopt the converse approach, by starting with an appropriate ansatz for the distribution function, and thereby solving for the potential--density pair, but this is less commonly used \citep{an05,ea05,li13,ng13,li14}. 

However, there are methods by which one can compute moments of the distribution function without actually determining the specific form of the distribution function. The most commonly used approach, when the underlying distribution function is isotropic, is to take advantage of the Jeans equations \citep{bi87}, and solve for the isotropic velocity dispersion. On the other hand, when the distribution function is anisotropic, the augmented density plays a significant role. For the case of spherical symmetry, it is a function of $r$ and the gravitational potential $\Phi$, and is obtained by integrating the distribution function over the velocities. If the augmented density has been obtained, one can use it directly to compute the velocity dispersions and the non-zero anisotropy parameter \citep{dj86,dj87,bvh07}. In this paper, we shall make use of these methods to derive the dispersion profiles and the anisotropic parameter through the construction of a suitable augmented density.

The primary aim of this paper is to study the effects of a cosmological constant on collisionless astrophysical systems at small scales. The cosmological constant $\lambda$ enters the Poisson equation when the Newtonian limit of general relativity \citep{no01,ax12,ax13} is taken, yielding
\begin{equation} \label{PoissonL}
\nabla^2 \Phi = 4\pi G \rho_M - \Lambda.
\end{equation}
If one associates $\Phi$ with the gravitational potential exerted by the matter, one can interpret $\lambda$ as the contribution arising from the presence of a repulsive harmonic potential \citep{no01,bal07,mer12}. The cosmological constant $\lambda$ has a long and varied history, see e.g. \citet{we08} and references therein. The value of the cosmological constant has been determined through several avenues \citep{we89,teg04}, which have yielded consistent results.  Although the effects of the cosmological constant on cosmological scales have been widely studied, its effects at smaller scales have not received the same level of attention. 

In recent times, the trend has begun to reverse, and the dynamical effects of the cosmological constant on gravitational and gravothermal instabilities \citep{ax00,ax12,ax13}, self-gravitating gases \citep{dev05a,dev05b}, general relativistic effects \citep{bo05a,bo05b} and astrophysical scales \citep{bal05,bal06,bal07}. A series of papers by Chernin and his collaborators \citep{te08,ch09,ch10,ch13} showed that the  cosmological constant played a significant role in the dynamics of galactic clusters. Some of the ensuing theoretical consequences were studied by \citet{bis12,mer12,bis13}. Very recently, the self-consistent approach that employed a Maxwell--Boltzmann distribution with a cutoff was studied by \citet{mer14} in the presence of a cosmological constant. 

However, a systematic study of the effects of dark energy on the observable properties such as the circular velocity, projected quantities (surface density, line-of-sight velocity dispersions), velocity dispersions, etc does not appear to have been undertaken. In this paper, we seek to redress this issue by carrying out such an analysis for a family of potential--density pairs which are widely used in the astrophysical literature. This family, first studied by \citet{ve79a,ve79b}, includes the \citet{pl11} and \citet{he90} models as special subcases. In particular, it was shown by \citet{ea05} that these models also possess a remarkably simple distribution function, with some additional special properties.

The outline of the paper is as follows. Some of the mathematical preliminaries used in the subsequent sections are presented in Section \ref{SecII}. In Section \ref{SecIII}, we construct the Veltmann models in the presence of a negative cosmological constant, and analyse the consequences. The case with a positive cosmological constant is studied in Section \ref{SecIV}. In Section \ref{SecIVPart2}, we study the effects of surrounding systems and compare them against those generated by the cosmological constant. Lastly, we conclude with Section \ref{SecV}, which summarizes our results. The details behind the calculations are presented in Appendix \ref{AppA}.

\section{The mathematical preliminaries} \label{SecII}
We briefly describe the Veltmann models and their significance, and develop some of the mathematical methods that we use in the subsequent sections. 

We use the self-consistent approach where a particular solution $f({\bf{r}},{\bf{v}},t)$ of the Vlasov equation (the collisionless Boltzmann equation) is used to obtain the density, which is then solved to determine the potential $\Phi$ via Poisson's equation. Our set of equations is
\begin{equation} \label{VPsystem}
\rho = \int f d^3{\bf{v}}, \quad \quad 4\pi G \rho = \nabla^2 \Phi.
\end{equation}
From Jeans' theorem, we know that the steady-state distribution function can be expressed via the integrals of motion. When one assumes spherical symmetry, the relevant integrals of motion are the energy (per unit mass) $E=v^2/2+\Phi$ and the magnitude of the angular momentum $L= \left|\bf{r} \times \bf{v}\right|$. However, the common convention is to work with the \emph{relative potential} and the \emph{relative energy} as defined in \citet{bi87}, which are given by
\begin{equation} \label{relativePotE}
\Psi \equiv -\Phi + \phi_c \quad {\mathrm{and}} \quad \cale \equiv -E + \phi_c = \Psi - \frac{1}{2}v^2,
\end{equation}
and $\phi_c$ is identified with the potential at the boundary of the system. It is evident that $\phi_c = 0$ for systems with infinite extent \citep{mo10}.

Let us consider the following set of one-parameter potential--density pairs
\begin{equation}\label{phiVeltmann}
\Phi=-\frac{GM}{\left(a^p+r^p\right)^{1/p}},
\end{equation}
\begin{equation}\label{rhoVeltmann}
\rho(r)=\frac{(1+p)M}{4\pi}\frac{a^p}{r^{2-p}\left(a^p+r^p\right)^{2+(1/p)}},
\end{equation}
where the parameter $p$ is positive. This family is a particular case of a larger group of models described in \citet{ve79a,ve79b}, and it is convenient henceforth to refer to them as the Veltmann models. They have been studied a great deal in the astrophysical literature since the Plummer ($p=2$) and the Hernquist ($p=1$) models are special cases of this family. In addition, it is known that the case $p=\frac{1}{2}$ is similar to the famous NFW profile used to model dark matter haloes \citep{zh96}. It was shown by \citet{ea05} that this family possessed a remarkably simple anisotropic distribution function, which was expressed as a double power function of $\mathcal{E}$ and $L$. The family of distribution functions obtained also satisfied the virial theorem locally, leading to them being labelled `hypervirial'. Double power models for other potential--density pairs that satisfied the property of hyperviriality were subsequently constructed by \citet{an05,li14}. 

After integrating $f$ w.r.t the velocities, we obtain $\rho{(r,\Phi)}$, which is known as the augmented density. It is extremely useful in computing higher order moments of the distribution function, without actually determining the distribution function itself. In particular, one can use it to compute the velocity dispersions following the approach outlined in \citet{dj87} which yields
\begin{equation} \label{sigmar2}
\sigma_r^2 \equiv \left\langle v_{r}^{2}\right\rangle  = -\frac{1}{\rho(\Phi,r)} \int_0^\Phi{\rho(\Phi',r)}d\Phi',
\end{equation}
\begin{equation} \label{sigmat2}
\sigma_\theta^2 \equiv \left\langle v_{\theta}^{2}\right\rangle = -\frac{1}{\rho(\Phi,r)} \int_{0}^{\Phi} \partial_{r^2}\left[r^2\rho(\Phi',r)\right] d\Phi',
\end{equation}
and it must be noted that $\sigma_\theta^2 = \sigma_\phi^2$. Similarly, we note that the expectation values of $v_r$, $v_\theta$ and $v_\phi$ are all zero as well. We refer the reader to \citet{li14} for a detailed discussion of the same. The anisotropy parameter $\beta$ is
\begin{equation} \label{betafromv}
\beta = 1 - \frac{\sigma_\theta^2}{\sigma_r^2}.
\end{equation}

\section{The Veltmann models with a negative cosmological constant} \label{SecIII}

\begin{figure*}
$$
\begin{array}{ccc}
 \includegraphics[width=5.2cm]{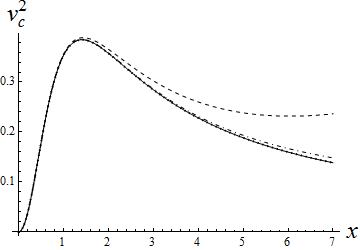} & \includegraphics[width=5.2cm]{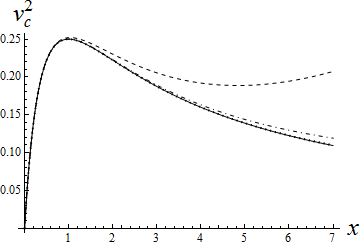} & \includegraphics[width=5.2cm]{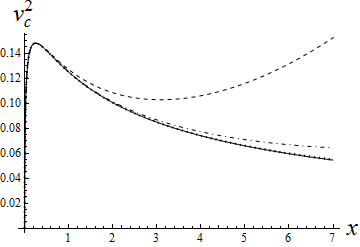}\\
 \quad\quad(a) & \quad\quad(b) & \quad\quad(c)\\
\end{array}
$$
\caption{The circular velocity as a function of $x$ for the $\lambda>0$ Veltmann models. Left-hand panel: circular velocity for the $p=2$ Plummer model. Central panel: circular velocity for the $p=1$ Hernquist model. Right-hand panel: circular velocity for the $p=1/2$ NFW-like model. For all panels: the curves with unbroken, thick-dotted, dotted, dot-dashed and dashed correspond to the values of $\epsilon=0,10^{-6},10^{-5},10^{-4},10^{-3}$ respectively. Note that the circular velocities have been converted to dimensionless units before plotting.}
\label{fig1}
\end{figure*}

\begin{figure*}
$$
\begin{array}{ccc}
 \includegraphics[width=5.2cm]{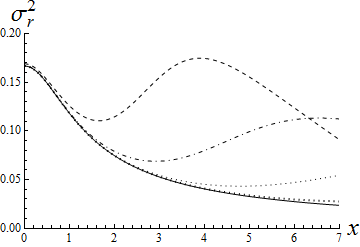} & \includegraphics[width=5.2cm]{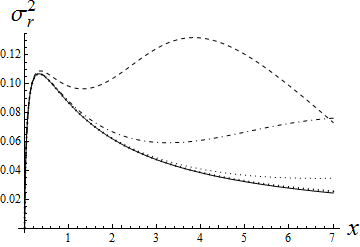} & \includegraphics[width=5.2cm]{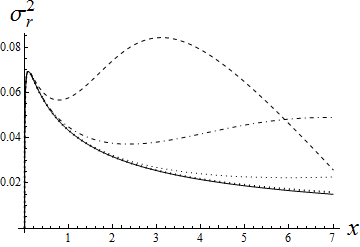}\\
 \quad\quad(a) & \quad\quad(b) & \quad\quad(c)\\
\end{array}
$$
\caption{The isotropic velocity dispersion as a function of $x$ for the $\lambda>0$ Veltmann models. Left-hand panel: isotropic velocity dispersion for the $p=2$ Plummer model. Central panel: isotropic velocity dispersion for the $p=1$ Hernquist model. Right-hand panel: isotropic velocity dispersion for the $p=1/2$ NFW-like model. For all panels: the curves with unbroken, thick-dotted, dotted, dot-dashed and dashed correspond to the values of $\epsilon=0,10^{-6},10^{-5},10^{-4},10^{-3}$ respectively. Note that the velocity dispersions have been converted to dimensionless units before plotting.}
\label{fig3}
\end{figure*}

In this section, we discuss the Veltmann models with an additional potential term included, that mimics the behaviour of a negative cosmological constant. We also discuss some of the implications on the observable properties, when such a term is included. Although the commonly accepted scenario involves the case with a positive cosmological constant, there have been attempts to investigate the signatures and effects of a negative cosmological constant in cosmology and astrophysics \citep{ca02,bal05,bal07,ca08,ha12}.

\subsection{The basic features}
Let us denote the potential of the Veltmann models by $\Phi_M$ where the label $M$ indicates it is the potential arising from the contribution of matter (baryonic and/or dark). To this, we add a new potential, labelled $\Phi_\lambda$ such that the total potential becomes
\begin{equation} \label{totalpot}
\Phi = -\frac{GM}{\left(a^p+r^p\right)^{1/p}} + \frac{1}{6} \lambda r^2 = \Phi_M + \Phi_\lambda,
\end{equation}
and the corresponding density is found via Poisson's equation to be
\begin{equation} \label{totalden}
4 \pi G \rho_M + \lambda =  \frac{G M (1+p) a^p}{r^{2-p}\left(a^p+r^p\right)^{2+(1/p)}} + \lambda = \nabla^2 \Phi.
\end{equation}
From the above expression, we see that this is identical to the Newtonian limit of the Einstein field equations, if one identifies $\lambda$ with the cosmological constant \citep{no01,ax12,ax13}. But, the crucial difference is that the above equation describes a \emph{negative} cosmological constant, since we take $\lambda>0$ in this section. In the next section, we assume $\lambda<0$, which amounts to replacing $\lambda$ by $-\lambda$. As a result, equation (\ref{totalden}) becomes
\begin{equation}
4 \pi G \rho_M - \lambda = \nabla^2 \Phi,
\end{equation}
which corresponds to the commonly studied case with a \emph{positive} cosmological constant.
Hence, we can use the potential (\ref{totalpot}) to model the effects of a negative cosmological constant on collisionless astrophysical systems. 

A few observations regarding the above equations are necessary before proceeding further. The presence of the additional term in (\ref{totalden}) leads to the condition that the mass of the system becomes infinite if the system is of infinite extent. Hence, this requires us to impose a cut-off radius. From (\ref{totalpot}), we see that $\Phi_M$ is attractive and $\Phi_\lambda$ is repulsive, as seen from their signs, provided that $\lambda >0$ -- an assumption we use throughout the rest of this section. This naturally suggest a cutoff radius where the two balance each other, i.e. we define $\xi$ such that 
\begin{equation}
\frac{GM}{\left(a^p+\xi^p\right)^{1/p}} = \frac{1}{6} \lambda \xi^2.
\end{equation}
This assumption also has an important consequence when it comes to computing the distribution function. In (\ref{relativePotE}), we introduced the variable $\phi_c$, the potential at the boundary. With our choice of cut-off radius, we obtain $\phi_c=\Phi\left(\xi\right) = 0$ and hence $\Psi = -\Phi$. In other words, this choice of cut-off radius effectively allows us to model our system in a manner similar to systems with infinite extent, since the property $\phi_c = 0$ is common to both.  Next, we note that if $\lambda = 0$, one finds that $\xi \rightarrow \infty$ and we recover the case with infinite extent. We introduce the variables
\begin{equation} \label{newvar}
\Phi_0 = \frac{GM}{a}, \quad \rho_0 = \frac{M}{4\pi a^3}, \quad x = \frac{r}{a}, \quad \epsilon = \frac{\lambda a^3}{6 G M},
\end{equation}
which allows us to express the potential--density pair as follows
\begin{equation} \label{dimlesspot}
\Psi = - \Phi = \Phi_0\left[\frac{1}{\left(1+x^p\right)^{1/p}} - \epsilon x^2\right],
\end{equation}
\begin{equation} \label{dimlessrho}
\rho = \rho_0 \left[\frac{(1+p)x^{p-2}}{\left(1+x^p\right)^{2+(1/p)}} + 6 \epsilon\right].
\end{equation}
The dimensionless cutoff radius, defined via $\chi=\xi/a$ satisfies $\Psi = 0$, where $\Psi$ is given by (\ref{dimlesspot}).

The most common method for determining distribution functions relies on Eddington's formula, which involves finding $\rho$ as a function of $\Psi$ and performing an Abel inversion. However, in order to do so we need to solve for $r(\Psi)$ and then plug into $\rho(r)$. From (\ref{dimlesspot}), it is clear that one would need to solve a complicated algebraic equation in $x$. For the Plummer and Hernquist models, one can convert the equation to a cubic in $\sqrt{1+x^2}$ and $x$, respectively. Despite this simplification,  the inversion that determines $\rho(\Psi)$ is quite involved. Instead, we use (\ref{dimlesspot}) to obtain
\begin{equation} \label{rhoxrel}
\frac{1}{\left(1+x^p\right)^{1/p}} = \frac{\Psi}{\Phi_0} + \epsilon x^2,
\end{equation}
and we shall introduce the augmented density
\begin{eqnarray} \label{rhoaug1}
\frac{\rho}{\rho_0} &=& \Bigg[ (1+p) \left(\frac{\Psi}{\Phi_0} + \epsilon x^2\right)^{2p+1} x^{p-2} \\ \nonumber
&& + \,\, 6\epsilon \left(1 + x^p\right)^\nu \left(\frac{\Psi}{\Phi_0} + \epsilon x^2\right)^{\nu p} \Bigg],
\end{eqnarray}
which was constructed by the substitution of (\ref{rhoxrel}) into (\ref{dimlessrho}) and some subsequent algebraic manipulations. In the above expression, the parameter $\nu$ is an arbitrary positive number. From the above expression, the familiar hypervirial family \citep{ea05} is recovered on setting $\epsilon = 0$. Furthermore, it is evident that the augmented density is a valid one since it is always positive. This follows from the fact that $x$, $\epsilon$ and $\Psi$ are strictly positive. One could easily make up increasingly complex augmented densities by multiplying and dividing with the factors introduced in (\ref{rhoxrel}), but this is not the primary aim of the paper. 

There are multiple methods that can be used to obtain the distribution function from the given augmented density. The first involves the application of the Laplace-Mellin formalism outlined in \citet{dj86} and \citet{bvh07}, and the references therein, by carrying out a complicated double-integral transform. Alternatively, we could make use of the fractional derivative approach as outlined in \citet{pe08b} or the equivalent approach developed in \citet{ji07a,ji07b,ji07c}. The augmented density is non-separable, which implies that the phase-space consistency requirements derived in \citet{an11a,an11b,an12} will not be directly applicable here. As a result, it is evident that computing the distribution function and analysing its validity via phase-space consistency checks are both complicated procedures. Hence, we shall instead work with the augmented density (\ref{rhoaug1}) which is known to be non-negative, and use it to compute the velocity dispersions.

We reiterate that the augmented density (\ref{rhoaug1}) is by no means unique, and although it is non-negative, there is no such guarantee for its corresponding distribution function. This augmented density was chosen since it reduces to the hypervirial family, which has been studied elsewhere in the literature, and because it possesses a relatively uncomplicated form.

\begin{figure*}
$$
\begin{array}{ccc}
 \includegraphics[width=5.2cm]{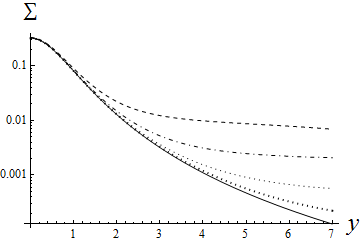} & \includegraphics[width=5.2cm]{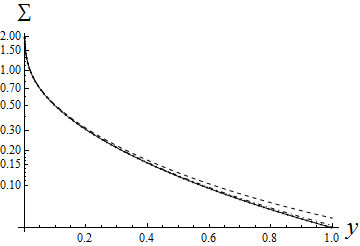} & \includegraphics[width=5.2cm]{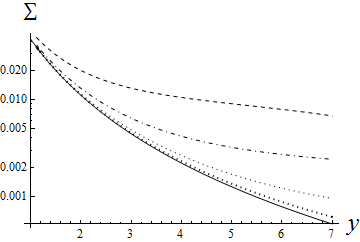}\\
 \quad\quad(a) & \quad\quad(b) & \quad\quad(c)\\
\end{array}
$$
\caption{The surface density as a function of $y=R/a$ for the $\lambda>0$ Veltmann models. Left-hand panel: surface density for the $p=2$ Plummer model. Central panel: surface density for the $p=1$ Hernquist model within the range $0<y<1$. Right-hand panel: surface density for the $p=1$ Hernquist model within the range $1<y<7$. For all panels: the curves with unbroken, thick-dotted, dotted, dot-dashed and dashed correspond to the values of $\epsilon=0,10^{-6},10^{-5},10^{-4},10^{-3}$ respectively. Note that the surface densities have been converted to dimensionless units before plotting.}
\label{fig5}
\end{figure*}

\begin{figure*}
$$
\begin{array}{ccc}
 \includegraphics[width=5.28cm]{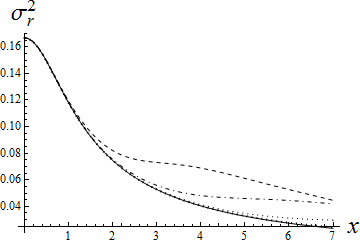} & \includegraphics[width=5.28cm]{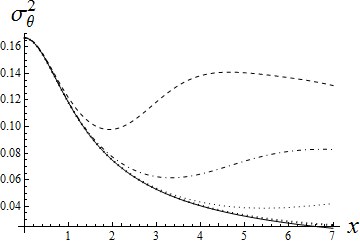} &   \includegraphics[width=5.28cm]{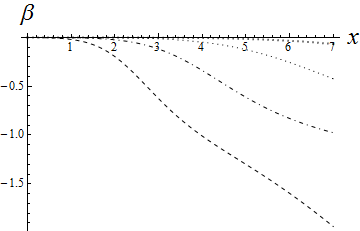}\\
 \quad\quad(a) & \quad\quad(b) &  \quad\quad(c)\\
\includegraphics[width=5.28cm]{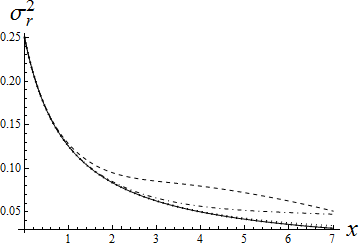} & \includegraphics[width=5.28cm]{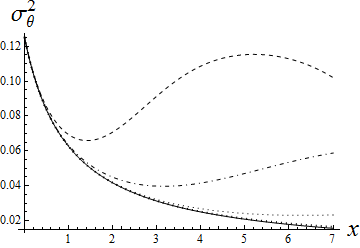} & \includegraphics[width=5.28cm]{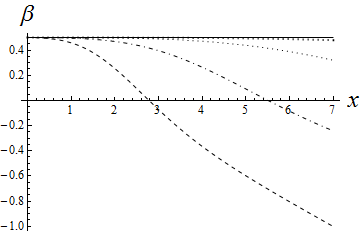}\\
 \quad\quad(d) & \quad\quad(e) & \quad\quad(f) \\
 \includegraphics[width=5.28cm]{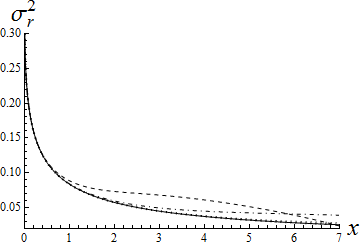} & \includegraphics[width=5.28cm]{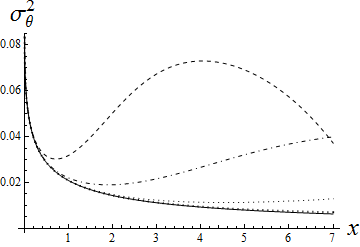} & \includegraphics[width=5.28cm]{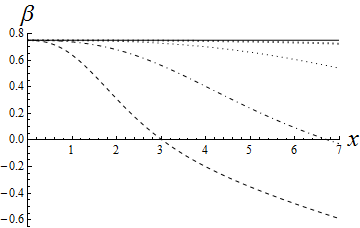}\\
 \quad\quad(g) & \quad\quad(h) & \quad\quad(i) \\
\end{array}
$$
\caption{All of the above panels depict the $\lambda>0$ models. Panels (a), (d) and (g) depict $\sigma_r^2$ vs $x$ for the Plummer, Hernquist and NFW-like profiles respectively. Panels (b), (e) and (h) depict $\sigma_\theta^2$ vs $x$ for the Plummer, Hernquist and NFW-like profiles respectively. Panels (c), (f) and (i) depict $\beta$ vs $x$ for the Plummer, Hernquist and NFW-like profiles respectively. For all panels: the curves with unbroken, thick-dotted, dotted, dot-dashed and dashed correspond to the values of $\epsilon=0,10^{-6},10^{-5},10^{-4},10^{-3}$ respectively; the velocity dispersions are given in dimensionless units.}
\label{fig7}
\end{figure*}

\subsection{Discussion}

\begin{figure*}
$$
\begin{array}{ccc}
 \includegraphics[width=5.2cm]{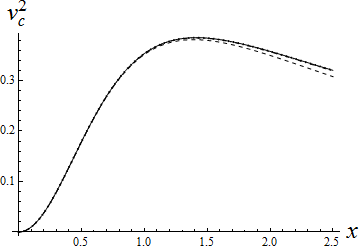} & \includegraphics[width=5.2cm]{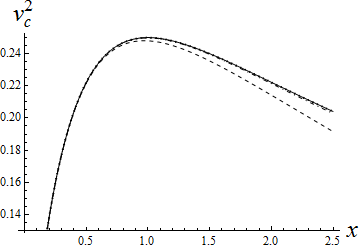} & \includegraphics[width=5.2cm]{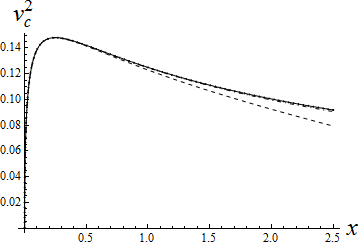}\\
 \quad\quad(a) & \quad\quad(b) & \quad\quad(c)\\
\end{array}
$$
\caption{The circular velocity as a function of $x$ for the $\lambda<0$ Veltmann models. Left-hand panel: circular velocity for the $p=2$ Plummer model. Central panel: circular velocity for the $p=1$ Hernquist model. Right-hand panel: circular velocity for the $p=1/2$ NFW-like model. For all panels: the curves with unbroken, thick-dotted, dotted, dot-dashed and dashed correspond to the values of $\epsilon=0,10^{-6},10^{-5},10^{-4},10^{-3}$ respectively. Note that the circular velocities have been converted to dimensionless units before plotting.}
\label{fig2}
\end{figure*}

\begin{figure*}
$$
\begin{array}{ccc}
 \includegraphics[width=5.2cm]{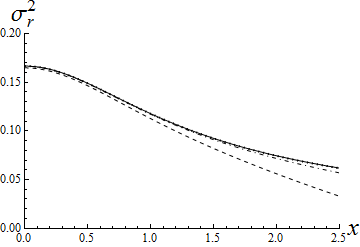} & \includegraphics[width=5.2cm]{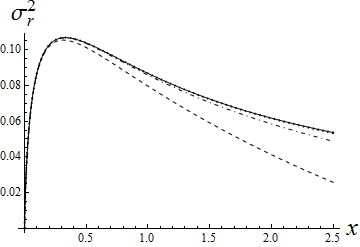} & \includegraphics[width=5.2cm]{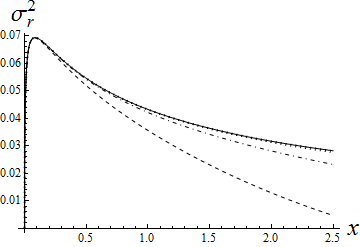}\\
 \quad\quad(a) & \quad\quad(b) & \quad\quad(c)\\
\end{array}
$$
\caption{The isotropic velocity dispersion as a function of $x$ for the $\lambda<0$ Veltmann models. Left-hand panel: isotropic velocity dispersion for the $p=2$ Plummer model. Central panel: isotropic velocity dispersion for the $p=1$ Hernquist model. Right-hand panel: isotropic velocity dispersion for the $p=1/2$ NFW-like model. For all panels: the curves with unbroken, thick-dotted, dotted, dot-dashed and dashed correspond to the values of $\epsilon=0,10^{-6},10^{-5},10^{-4},10^{-3}$ respectively. Note that the velocity dispersions have been converted to dimensionless units before plotting.}
\label{fig4}
\end{figure*}

We return to (\ref{newvar}) where we first introduced the value of $\epsilon$. In order to proceed further, we must obtain an estimate as to the value of $\epsilon$. Because the cosmological constant is not expected to play a significant role (in general) at small scales, we expect $\epsilon$ to be quite small. For our system, we know that $a$ and $M$ represent the characteristic length scale and total mass of the matter component. Hence, we can identify the factor of $M/a^3$ that appears in (\ref{newvar}) with the characteristic mass density of the matter component, which we denote by $\rho_m$. We see that $\epsilon \propto \rho_m^{-1}$, holding $\lambda$ fixed, and hence the contributions of $\epsilon$ to the overall dynamics become important when $\rho_m$ is low. Such a situation is more commonly encountered in dark matter haloes, which are much more rarefied than their baryonic counterparts (galaxies). 

For instance, if we consider our Milky Way halo \citep{bat05} and assume $M\sim 10^{12} M_{\odot}$ and $a\sim 120\,\mathrm{kpc}$, we find that the value of $\epsilon$ is approximately in the range $10^{-3}-10^{-4}$. In the solar neighbourhood \citep{bah84}, if we assume that the density is roughly $0.1 M_\odot \, \mathrm{pc}^{-3}$, we obtain $\epsilon \sim \calo\left(10^{-5}\right)$. If we consider the regions in the galactic centre, the value of $\rho_m$ is even higher, and that of $\epsilon$ is consequently lowered. In Table. \ref{Tab1}, some of the characteristic values of $\epsilon$ and their cut-off radii are presented. 

The total mass present in this system, up to the cutoff radius is found by integrating (\ref{dimlessrho}) from $x=0$ to $x=\chi$ and is found to be
\begin{equation} \label{MassMPos}
\frac{M_\chi}{M} = \left(\epsilon \chi^3\right) \left[2+ \left(\epsilon \chi^3\right)^p\right].
\end{equation}  The ratio of $M_\chi$ to $M$ can be interpreted as the ratio of the total mass arising from the contribution of dark energy + matter versus the total mass present if there were zero dark energy. The choice of $\chi$ was motivated by choosing the point where the potentials cancelled out, but it must be emphasized that the cutoff radius is totally arbitrary. Hence, this ratio is not very significant, although the situation is very different for $\lambda < 0$, as we shall see. However, one important point does deserve to be mentioned. For $p=1$ and $p=2$, one finds that $\epsilon \chi^3 \approx 1$, which implies that $\frac{M_\chi}{M} \approx 3$. The result is (nearly) independent of the choice of $p$ and $\epsilon$.

\begin{table}
\caption{$\chi$ values for Veltmann models}
\label{Tab1}
\begin{tabular}{|c|c|c|c|}
\hline 
$\epsilon$ & $\chi(p=1/2)$ & $\chi(p=1)$ & $\chi\left(p=2\right)$\tabularnewline
\hline 
\hline 
$10^{-3}$ & $8.19$ & $9.68$ & $9.98$\tabularnewline
\hline 
$10^{-4}$ & $18.78$ & $21.2$ & $21.54$\tabularnewline
\hline 
$10^{-5}$ & $42.19$ & $46.1$ & $46.41$\tabularnewline
\hline 
$10^{-6}$ & $93.65$ & $99.7$ & $100$\tabularnewline
\hline 
\end{tabular}
\medskip

The table illustrates how the same values of $\epsilon$ lead to (minor) differences in the value of $\chi$ for different models. The value of $\chi$ is computed for the NFW-like ($p=1/2$), Hernquist ($p=1$) and Plummer($p=2$) models, for several values of $\epsilon$.
\end{table}

We shall now discuss some of the effects on the observable properties of the Veltmann models that arise from a non-zero $\lambda$ (or equivalently $\epsilon$). The calculations have been delegated to Appendix \ref{AppA}, and their final implications are discussed in this subsection.

In Fig. \ref{fig1}, the circular velocities for the Plummer, Hernquist and NFW-like models are depicted. It is seen that the case with $\epsilon = 10^{-3}$ changes the curves significantly, which implies that the effects of $\lambda$ become significant in dilute environments. In particular, note that the circular velocities fall off as $r^{-1}$ (Keplerian behaviour) for large values of $r$, and this is counteracted by the $\epsilon x^2$ term present in (\ref{circvelpos}). 

A much more remarkable effect arising from the addition of $\lambda$ is to be found in the isotropic velocity dispersion profiles for the Plummer, Hernquist and NFW-like models. In Fig. \ref{fig3}, it is seen that the cases with $\epsilon=10^{-3}$ and $\epsilon=10^{-4}$ alter the profiles significantly, particularly the former. In fact, we see that the profiles become non-monotonic for all non-zero values of $\epsilon$, with the possible exception of the cases where $\epsilon=10^{-6}$. It is evident that curves with differing values of $\epsilon$ give rise to visibly different profiles. Hence, one could potentially use the structure of the curves to either constrain the characteristic density $\rho_m$ or the value of $\lambda$ since $\epsilon \propto \lambda \left(\rho_m\right)^{-1}$. 

The dimensionless surface density profiles are plotted as a function of $y=R/a$ in Fig. \ref{fig5}. The NFW-like profile is not depicted, since it cannot be easily expressed in terms of elementary functions, and was consequently not calculated in Appendix \ref{AppA}. The surface density for the Hernquist model splits into two different algebraic expressions, one for $y<1$ and the other for $y>1$. However, it is continuous and differentiable across the transition point $y=1$. As a result, there are two plots for the Hernquist surface density in Fig. \ref{fig5}, to account for the different expressions for $\Sigma$. From Fig. \ref{fig5}, we see that the surface density increases when $\epsilon$ is higher, and the density profiles with different $\epsilon$-values are easily distinguishable for higher values of $y$.

For the augmented density introduced in (\ref{rhoaug1}), the corresponding velocity dispersions and the anisotropy parameter were computed in Appendix \ref{AppA}. These profiles are plotted in Fig. \ref{fig7} where $\nu = 1$ was chosen for all the panels. It is seen that higher values of $\epsilon$ lead to significant changes in the structure of the dispersion profiles. The anisotropy parameter $\beta$ is even more sensitive to distinguishing between profiles with different values of $\epsilon$ in the outer regions. It is seen that some of the velocity dispersion profiles for higher values of $\epsilon$ are highly non-monotonic. For $p<2$, the density has a cusp around the origin, and we compute the slope of the cusp, denoted by $\gamma_0$, and the anisotropy parameter at $x=0$, denoted by $\beta_0$. We find that the cusp slope--central anisotropy inequality \citep{ae06} is saturated, i.e. we have $\gamma_0 = 2\beta_0$ for the distribution functions arising from the augmented density (\ref{rhoaug1}). 

\section{The Veltmann models with a positive cosmological constant} \label{SecIV}

\begin{figure*}
$$
\begin{array}{ccc}
 \includegraphics[width=5.2cm]{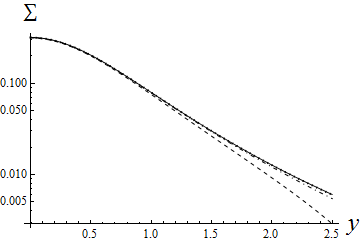} & \includegraphics[width=5.2cm]{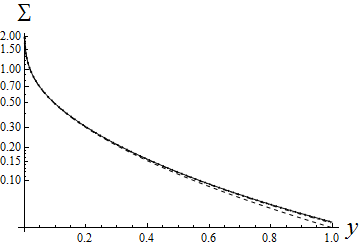} & \includegraphics[width=5.2cm]{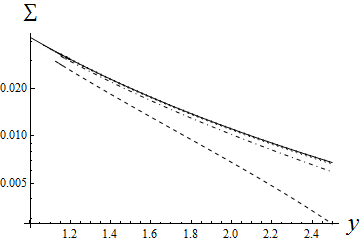}\\
 \quad\quad(a) & \quad\quad(b) & \quad\quad(c)\\
\end{array}
$$
\caption{The surface density as a function of $y=R/a$ for the $\lambda>0$ Veltmann models. Left-hand panel: surface density for the $p=2$ Plummer model. Central panel: surface density for the $p=1$ Hernquist model within the range $0<y<1$. Right-hand panel: surface density for the $p=1$ Hernquist model within the range $1<y<7$. For all panels: the curves with unbroken, thick-dotted, dotted, dot-dashed and dashed correspond to the values of $\epsilon=0,10^{-6},10^{-5},10^{-4},10^{-3}$ respectively. Note that the surface densities have been converted to dimensionless units before plotting.}
\label{fig6}
\end{figure*}

\begin{table*}
\begin{minipage}{126mm}
\caption{$\chi_d$ and $M_d$ values for Veltmann models}
\label{Tab2}
\begin{tabular}{|c|c|c|c|c|c|c|}
\hline 
$\epsilon$ & $\chi_{d}\left(p=1/2\right)$ & $\chi_{d}\left(p=1\right)$ & $\chi_{d}\left(p=2\right)$ & $M_{d}/M\left(p=1/2\right)$ & $M_{d}/M\left(p=1\right)$ & $M_{d}/M\left(p=2\right)$\tabularnewline
\hline 
\hline 
$10^{-3}$ & $2.84$ & $3.55$ & $3.32$ & $0.20$ & $0.52$ & $0.80$\tabularnewline
\hline 
$10^{-4}$ & $6.40$ & $6.86$ & $5.40$ & $0.32$ & $0.70$ & $0.92$\tabularnewline
\hline 
$10^{-5}$ & $13.75$ & $12.77$ & $8.65$ & $0.44$ & $0.82$ & $0.97$\tabularnewline
\hline 
$10^{-6}$ & $28.65$ & $23.28$ & $13.76$ & $0.55$ & $0.89$ & $0.99$\tabularnewline
\hline 
\end{tabular}
\medskip

The table illustrates how the same values of $\epsilon$ lead to (significant) differences in the values of $\chi_d$ and $M_d$ for different models. The value of $\chi_d$ and $M_d$ is computed for the NFW-like ($p=1/2$), Hernquist ($p=1$) and Plummer($p=2$) models, for differing values of $\epsilon$.
\end{minipage}
\end{table*}

In the previous section, we have assumed that $\lambda>0$, and we now consider the opposite scenario where $\lambda<0$. It is possible, simply via $\lambda \rightarrow -\lambda$, to study the dynamics in the presence of a positive cosmological constant.

This amounts to $\epsilon \rightarrow -\epsilon$ in most of the algebraic expressions obtained. Note that $\lambda$ and $\epsilon$ are still positive, but the additional negative sign allows us to model the system as if it were in the presence of a positive cosmological constant. The potential--density pair for these models is now given by
\begin{equation} \label{dimlesspotneg}
\Phi = -\Phi_0\left[\frac{1}{\left(1+x^p\right)^{1/p}} + \epsilon x^2\right],
\end{equation}
\begin{equation} \label{dimlessrhoneg}
\rho = \rho_0 \left[\frac{(1+p)x^{p-2}}{\left(1+x^p\right)^{2+(1/p)}} - 6 \epsilon\right].
\end{equation}
There are interesting and subtle differences between the models. The first arises from the difference in the cutoff radius. For the models in the previous section, the cutoff radius was arbitrary, and this degree of freedom enabled us to choose the cutoff radius $\chi$ such that $\Phi\left(x=\chi\right)=0$. In turn, this choice of $\chi$ led to the relation $\Psi = -\Phi$ through the use of (\ref{relativePotE}). In our case however, there is no arbitrariness possible in the choice of the `cut-off' radius. There are two radii that define themselves automatically, through the constraints of the system. 

The first is the point at which the density changes sign, and if we require the density to be always non-negative, this immediately constrains the maximum extent of the system. Let us denote this cutoff radius by $\chi_d$, which is found via setting (\ref{dimlessrhoneg}) to zero. 
Next, we consider the expression for the gravitational force. The force is proportional to $d\Phi/dx$, which is found to be
\begin{equation}
\frac{d\Phi}{dx} = \Phi_0 \left[\frac{x^{p-1}}{\left(1+x^p\right)^{1+(1/p)}} - 2\epsilon x\right],
\end{equation}
and this immediately suggests a second cutoff, when the above expression vanishes. Physically, this represents the point of zero gravity, where the force exerted by the matter is balanced by the contribution arising from the dark energy. We shall denote this second cutoff by $\chi_f$. We find that $\Phi\left(x=\chi_d\right)\neq0$ and $\Phi\left(x=\chi_f\right)\neq0$, which implies that we cannot simply set $\Psi = -\Phi$. The boundary contribution to the gravitational potential, evident from (\ref{relativePotE}), must also be included. Hence, we cannot simply replace $\epsilon \rightarrow -\epsilon$ in (\ref{rhoaug1}), since it was derived under the assumption that $\Psi = -\Phi$. Most of the mathematical expressions derived in Appendix \ref{AppA} are still valid, upon using $\epsilon \rightarrow -\epsilon$, since they do not presuppose any knowledge beyond the basic potential--density pair given by (\ref{dimlesspotneg}) and (\ref{dimlessrhoneg}).

We can compute the mass enclosed within the two different cutoff radii. If one uses $\chi_f$, it is clear that the mass enclosed is zero. This arises from the fact that $\frac{d\Phi}{dx} \propto v_c^2 \propto M(x)$, and since $\frac{d\Phi}{dx}=0$ at $x=\chi_f$, the mass enclosed is also zero. The second cutoff radius yields a mass
\begin{equation} \label{Massneglambda}
\frac{M_{d}}{M} = \frac{\chi_d^{p+1}}{\left(1+\chi_d^p\right)^{1+(1/p)}} - 2\epsilon \chi_d^3.
\end{equation}

We can obtain a much simpler estimate by using $\epsilon \ll 1$ and $x \gg 1$. The latter is not always true, even when the former is valid, but it suffices for a crude estimate. Using these approximations in (\ref{dimlessrhoneg}), we can solve for $\chi_d$, and then substitute the resultant value into (\ref{Massneglambda}). Thus, we find that
\begin{equation}
\frac{M_{d}}{M} \approx 1 - \left[\left(\frac{1+p}{3}\right)^3 (2\epsilon)^p \right]^{1/(p+3)},
\end{equation}
which implies that $\Delta M = M - M_{d} \sim M \epsilon^{p/(p+3)}$. 

In Table \ref{Tab2}, the cutoff radii $\chi_d$ and the corresponding masses $M_d$ are calculated for the NFW-like, Hernquist and Plummer models, for different values of $\epsilon$. For the Plummer model, we rapidly approach the value of near-unity for $M_d/M$ as one moves to lower values of $\epsilon$, but the situation is quite different for the Hernquist and NFW-like models, especially the latter. 

From Fig. \ref{fig2}, we see that the presence of a finite $\epsilon$ alters the circular velocity, particularly for the case where $\epsilon = 10^{-3}$. The effect experienced is exactly opposite to that seen in Fig. \ref{fig1}, since the rotation curves grow steeper, falling-off more rapidly as one moves to higher values of $\epsilon$. 

In Fig. \ref{fig4}, the isotropic velocity dispersions for the $\lambda<0$ cases are plotted for the three Veltmann models of interest. As one would expect, the plots obtained are quite different from the $\lambda>0$ counterparts plotted in Fig. \ref{fig3} and discussed in the previous section. As the value of $\epsilon$ is increased, these effects become more prominent, thereby enabling us to distinguish clearly between profiles with different values of $\epsilon$. The effects are opposite to those observed in the case where $\lambda>0$. This effect is also seen with respect to Fig. \ref{fig6}, as compared to Fig. \ref{fig5} (with $\lambda>0$). There are two plots for the Hernquist profile in Fig. \ref{fig6} since there exist two different algebraic expressions for $\Sigma$, as noted earlier. It is seen that the surface density is lowered as $\epsilon$ increases, and the profiles are increasingly distinguishable when one moves towards the outer regions of $y$.

\section{The gravitational effects from neighbouring objects} \label{SecIVPart2}

\begin{figure*}
$$
\begin{array}{ccc}
 \includegraphics[width=5.28cm]{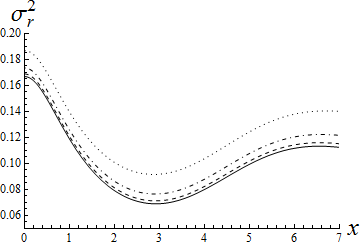} & \includegraphics[width=5.28cm]{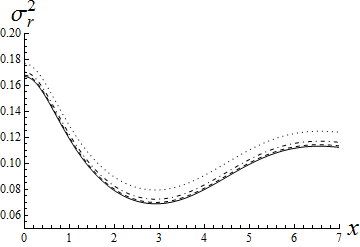} &   \includegraphics[width=5.28cm]{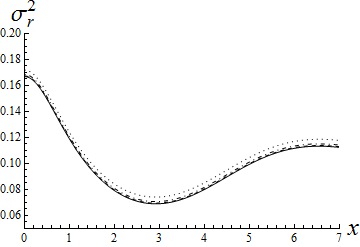}\\
 \quad\quad(a) & \quad\quad(b) &  \quad\quad(c)\\
\includegraphics[width=5.28cm]{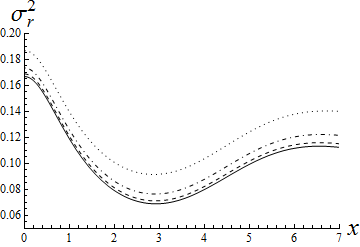} & \includegraphics[width=5.28cm]{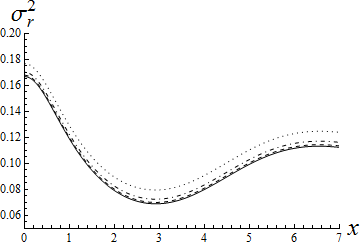} & \includegraphics[width=5.28cm]{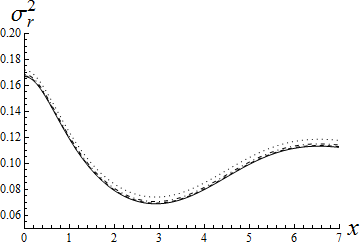}\\
 \quad\quad(d) & \quad\quad(e) & \quad\quad(f) \\
 \includegraphics[width=5.28cm]{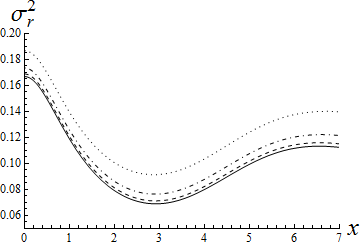} & \includegraphics[width=5.28cm]{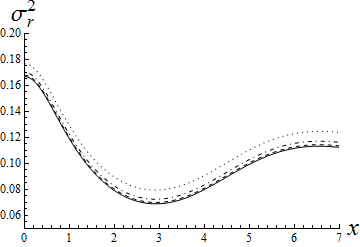} & \includegraphics[width=5.28cm]{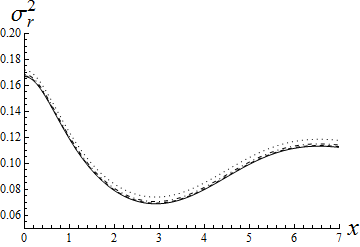}\\
 \quad\quad(g) & \quad\quad(h) & \quad\quad(i) \\
\end{array}
$$
\caption{All of the above panels depict the $\lambda>0$ Plummer models with $\epsilon=10^{-4}$. Panels (a), (b) and (c) have $\delta=0.5$ with $x_0=25,50,100$ respectively. Panels (d), (e) and (f) have $\delta=1$ with $x_0=25,50,100$ respectively. Panels (g), (h) and (i) have $\delta=3$ with $x_0=25,50,100$ respectively. For all panels: the curves with unbroken, dashed, dot-dashed and dotted correspond to the values of $\Delta=0,0.3,1,3$ respectively; the velocity dispersions are given in dimensionless units.}
\label{fig08}
\end{figure*}

From the preceding sections, we have seen that the effects of the
cosmological constant become discernible when $x\sim{\cal O}\left(1\right)$,
i.e. in regions where the radial distance $r$ is a few times larger
than the scale length $a$. However, it must be noted that other effects
come into play in these regions as well - the gravitational effects
of other (external) objects become important; by `objects' we
are referring to the type of system under consideration such as
galaxies, globular clusters, etc. As a result, it is likely that the
additional gravitational effects may wash out the effects of the cosmological
constant, which would prevent any experimental verification of these
effects. In order to analyse such a scenario, we consider the following
toy model to better understand the situation.

Let us call our main object as the `parent' and the far-off neighbour
as the `perturber'. We assume that the distance between the two
is denoted by $R_{0}$. As a result, we can use the law of cosines
to conclude that the distance of the perturber from the radial observation
point $r$ is given by $r'=\sqrt{R_{0}^{2}+r^{2}-2R_{0}r\cos\gamma}$.
As we have modelled the parent via the Veltmann models, we shall consider
the perturber as being described by a Veltmann model as well. Let
us denote the scale length of the perturber by $b$ and the mass by
$M_{p}$. Now, we introduce the following assumptions:
\begin{enumerate}
\item $R_{0}\gg a$ and $R_{0}\gg b$ can be justified since the inter-distance
between the two objects is much higher than the scale length of the
object. This is mostly the case in many astrophysical systems.
\item $R_{0}\gg r$ is justified on the grounds that we are interested
in values of $r$ which are typically of the order of the scale length $a$ 
(a few times higher).
\end{enumerate}
As a result, note that $r'$, which ranges from $R_{0}-r$ to $R_{0}+r$
is very nearly equal to $R_{0}$ through the application of the second
assumption. For the sake of simplicity henceforth, we shall assume
that $\gamma=0$, i.e. the parent, the perturber and the radial observation
point all lie on a straight line implying that $r'=R_{0}-r$. The
Veltmann potential-density pair for the perturber is found by noting
that it identical to that of (\ref{dimlesspot}) and (\ref{dimlessrho}), with $r$ replaced
by $r'$, the absence of the term involving $\epsilon$, and the other parameters suitably changed as well. As a result, we obtain
\begin{eqnarray}
\Psi_{\mathrm{ext}}&=&-\Phi_{\mathrm{ext}}=\Phi_{0}\frac{\Delta}{\left(\delta^{m}+x'^{m}\right)^{1/m}}, \nonumber \\
\rho_{\mathrm{ext}}&=&\rho_{0}\frac{\left(1+m\right)\Delta\delta^{m}x'^{m-2}}{\left(\delta^{m}+x'^{m}\right)^{2+(1/m)}},
\end{eqnarray}
where the new parameters are defined to be
\begin{equation}
\Delta=\frac{M_{p}}{M},\quad\delta=\frac{b}{a},\quad x'=\frac{R_{0}}{a}-x=x_{0}-x.
\end{equation}
Before proceeding further, let us carry out an estimate as to how
the density of the perturber stacks up against the parent density
at a distance of $x\sim5$. From the two assumptions listed above,
we find that 
\begin{equation}
\rho_{\mathrm{ext}}\approx\rho_{0}\frac{\left(1+m\right)\Delta\delta^{m}}{\left(x_{0}-x\right)^{m+3}},
\end{equation}
and the density arising from the cosmological constant is given by
\begin{equation}
\rho_{\lambda}=\rho_{0}\left(1+p\right)\left[\frac{6\epsilon}{1+p}\right].
\end{equation}
Let us evaluate the ratio of the above two quantities by using the
values of $\Delta=1$, $\delta=2$ and $x_{0}=50$ in the above expressions.
These choices are motivated by our choice of the Milky Way (parent)
and M31 (perturber) system. In addition, we consider the case where
$p=m=1$ and $\epsilon=10^{-4}$, and we find that $\rho_{\mathrm{ext}}/\rho{}_{\lambda}\approx1.6\times10^{-3}$
indicating that the influence of the perturber in this particular
scenario is weak.

However, we are much more interested in the effects that the perturber
has on the physically relevant quantities, such as the velocity dispersion,
surface density, etc. A comprehensive study is beyond the scope of
our present work. A rigorous calculation of the (isotropic) velocity
dispersion would entail the use of $\rho_{\mathrm{tot}}=\rho+\rho_{\mathrm{ext}}$
in (\ref{isoJeans}), where $\rho$ is given by (\ref{dimlessrho}). Similarly, one would need to
define $\Phi_{\mathrm{tot}}=\Phi+\Phi_{\mathrm{ext}}$ with $\Phi$
given by (\ref{dimlesspot}) and use $\Phi_{\mathrm{tot}}$ in (\ref{isoJeans}). 

Instead, we present a semi-heuristic calculation. If we treat the
perturbed potential--density pair as an independent quantity, we can
find its isotropic velocity dispersion by using the appropriate Jeans'
equation in terms of $\rho_{\mathrm{ext}}$ and $\Phi_{\mathrm{ext}}$
alone. For the Plummer model, the calculation is simple \citep{dj87,ea05}, with the result
\begin{equation} \label{isovelpert}
\left(\sigma_{r}^{2}\right)_{\mathrm{ext}}=\frac{\Psi_{\mathrm{ext}}}{6}=\frac{\Phi_{0}}{6}\frac{\Delta}{\left(\delta^{2}+x'^{2}\right)^{1/2}}.
\end{equation}
Now, we can add this expression to the one obtained in (\ref{Plumisovel}), and plot the resultant velocity dispersion as a function of $x$, which is undertaken in Fig. \ref{fig08}. Note that Fig. \ref{fig08} depicts the effects of a gravitational perturber on the parent system with a negative cosmological constant; similar results to the ones described below are found even when the case with a positive cosmological constant are taken into account.

In all the panels, the case with $\Delta = 0$ corresponds to the absence of the perturber's effects, i.e. it amounts to the expression (\ref{isovelpert}) becoming identically zero. We find that increasing $\Delta$, which amounts to the perturber becoming increasingly massive, results in the velocity dispersion profiles becoming shifted upwards. In other words, the more massive the perturber, the greater its effects on the profiles, which is an expected result. Similarly, we see that the choices with lower $x_0$ lead to slightly higher values. This indicates that the closer the perturber to the main system, the greater its influence -- a result along expected lines. We also note that the dependence of the velocity profiles on $\delta$ is minimal for our choice of values for $\delta$. It is worth emphasizing that each of the panels depicts the same phenomenon - the  `structure' of the profiles is mostly left unchanged. In the previous sections, we showed that the cosmological constant altered the structure of the profiles significantly, when $\epsilon$ was suitably low; this is illustrated through Figs. \ref{fig3} and \ref{fig7}. Fig. \ref{fig08} suggests that this structure-altering feature of the cosmological constant is still preserved under the influence of external systems, as long as they are not massive or closely spaced enough. 

It is worth mentioning once more that the resultant velocity dispersion is \emph{not} accurate,
since it wasn't derived in a fully self-consistent manner by solving the radial 
component of Jeans' equations (\ref{isoJeans}), while using  
$\Phi_{\mathrm{tot}}$ and $\rho_{\mathrm{tot}}$ as the potential--density pair. Yet, it
does present us with a crude estimate of the gravitational effects
of the perturber on the Veltmann models with a non-zero cosmological
constant. In summary, if one considers a self-consistent system, there
are two additional effects that compete against one another to affect
the physically observable profiles. They correspond to the following. 
\begin{enumerate}
\item The cosmological constant, which becomes relevant on increasingly
larger scales (the outer regions) since its corresponding gravitational potential
increases as the square of the distance. 
\item The presence of neighbouring systems. In the outer regions, we move
closer to the neighbouring system, increasing its gravitational effects.
In turn, this influences the observable properties predicted by our
models.
\end{enumerate}
Let us determine the critical point $x_b$ where the density contributed
by the gravitational constant becomes comparable to that of the perturber.
This occurs when
\begin{equation} \label{balanceDens}
6\epsilon=\frac{\left(1+m\right)\Delta\delta^{m}x'^{m-2}}{\left(\delta^{m}+x'^{m}\right)^{2+(1/m)}}
\end{equation}
In Table \ref{Tab3}, the critical points have been determined for some of the
more commonly used Veltmann models, with suitable choices of the free
parameters. It is seen that the density from the neighbouring system
dominates the cosmological constant only at very large distances --
far outside the range of $x$ which we have plotted in all the previous
figures. 

\begin{table}
\caption{$x_b$ values for the Veltmann models}
\label{Tab3}
\begin{tabular}{|c|c|c|c|}
\hline 
$\epsilon$ & $x_{b}$ $\left(m=1/2\right)$ & $x_{b}$ $\left(m=1\right)$ & $x_{b}$ $\left(m=2\right)$\tabularnewline
\hline 
\hline 
$10^{-3}$ & $47.16$ & $46.45$ & $46.68$\tabularnewline
\hline 
$10^{-4}$ & $43.61$ & $43.14$ & $44.6$\tabularnewline
\hline 
$10^{-5}$ & $36.26$ & $37.20$ & $41.35$\tabularnewline
\hline 
$10^{-6}$ & $21.34$ & $26.73$ & $36.24$\tabularnewline
\hline 
\end{tabular}
\medskip

The table illustrates the dependence of the critical point $x_b$ on $\epsilon$ for different models. The value of $x_b$ is computed for the NFW-like ($p=1/2$), Hernquist ($p=1$) and Plummer($p=2$) models, for several values of $\epsilon$. In carrying out these computations using (\ref{balanceDens}), the values $\Delta = \delta = 1$ and $x_0 = 50$ were used throughout.
\end{table}

Before rounding off this section, we emphasize once more that several
features of our toy model represent simplifications -- the presence
of a single perturber, the assumption that the perturber lies along
the radial direction, and the calculation of the velocity dispersion
by breaking it down into the contribution from the parent system and
that of the perturber. However, many of these calculations do appear
to suggest that the velocity profiles, densities, etc. are not significantly
affected by the presence of neighbouring systems. It must be noted
that there are several instances where these effects may play a significant
role. If we suppose our parent system to be one of the Milky Way's
dwarf satellites, and the perturber to be the Milky Way, the huge
discrepancies in the masses result in a very high value of $\Delta$,
and this causes the velocity dispersions and net densities to be altered
significantly. A second such scenario is possible when the parent
and the perturber are relatively close, i.e. the value of $x_{0}$
is not very high, thereby leading to a higher degree of interaction.
In both these cases, we expect the effects of the cosmological constant
to be washed out. 

\section{Conclusion} \label{SecV}
In this paper, we have investigated the effects of an additional potential $\Phi_\lambda$ that mimics the behaviour of the cosmological constant. The case with $\lambda>0$ and $\lambda<0$ were both considered, and they yield contrasting results, as expected. In both cases, we computed the cut-off radii, and showed that the latter scenario gave rise to intrinsic cut-off radii. The mass present within this cut-off radius are also computed in both the cases, and it is shown that the cases with $\lambda<0$ (the positive cosmological constant) lead to model-dependent results. We show that quantities such as the circular velocity and the surface density are shifted upwards (downwards) by the presence of a negative (positive) cosmological constant, as expected. However, it was shown that each of these profiles became quite different as one moved towards higher values of the radius, allowing us to differentiate between the profiles.

The most remarkable effects of a finite value of $\epsilon$, defined via (\ref{newvar}), are exhibited through the velocity dispersion profiles. The presence of a negative $\epsilon$ leads to a steeper fall-off of the profiles, implying that the positive cosmological constant can play a role in altering the shape and structure of the profiles. However, the case where $\epsilon>0$ (negative cosmological constant) yields some unusual results. It is shown that the well-behaved dispersion profiles for $\epsilon=0$ grow increasingly non-monotonic for higher values of $\epsilon$. The anisotropy parameter is also very sensitive to these changes, even allowing us to distinguish between the smallest non-zero $\epsilon$ and the $\epsilon=0$ profiles. In other words, the studies of models with a negative cosmological constant are severely constrained by this analysis, since the structure of the velocity dispersion profiles is highly sensitive to the presence of a finite $\epsilon$.

The relation $\epsilon \propto \lambda \left(\rho_m\right)^{-1}$, where $\rho_m$ is the characteristic density of the matter component, also has important consequences. In dilute environments, we expect the corrections to the observable quantities, arising from the presence of a non-zero $\epsilon$, to become significant. Dark matter haloes represent one such possible environment, and it is possible that the effects of the cosmological constant may need to be taken into account when modelling them.

There remains one other possibility since the cosmological constant is shown to have discernible effects in the outer regions -- the influence of neighbouring systems. In Section \ref{SecIVPart2}, we presented a heuristic analysis of this scenario, and the conditions under which the neighbouring systems play a significant role were elucidated. It is suggested that the cosmological constant may still play an important role, except when the neighbours are closely spaced, or when they are much more massive than the central system. It is also suggested that the general shapes and structures of the velocity dispersion profiles are mostly preserved, even when neighbouring systems are taken into account. As a result, the structure-altering properties of the cosmological constant still appear to be present, even if diminished somewhat.

Avenues for future work include carrying out the computations to determine the rest of the projected quantities, and the construction of an explicit (and valid) distribution function for the Veltmann models in the presence of a finite cosmological constant. 

\section{ACKNOWLEDGEMENTS}
The author thanks Philip Morrison for his support and guidance. In addition, the comments and suggestions of the anonymous referee were very insightful, and greatly appreciated. This work was supported by U.S.\ Department of Energy Contract No. DE-FG05-80ET-53088.


\appendix
\section{Observable properties for the $\lambda>0$ models} \label{AppA}
 
The circular velocity for the Veltmann models in the presence of a non-zero $\lambda$ is given by
\begin{equation} \label{circvelpos}
v_c^2 = x \frac{d\Phi}{dx} = \Phi_0 \left[\frac{x^p}{\left(1+x^p\right)^{1+(1/p)}} + 2\epsilon x^2\right].
\end{equation}

One can also compute the surface density via
\begin{equation}
\Sigma = 2 \int_R^\infty \frac{\rho(r)r\,dr}{\sqrt{r^2-R^2}},
\end{equation}
and the expressions turn out to be rather complicated for a generic value of $p$. In the above expression, one must note that the upper bound of $r \rightarrow \infty$ must be replaced by the cutoff radius, since our system is of finite extent. For the Plummer model, we have
\begin{equation} \label{Plumsurf}
\tilde{\Sigma}(y)=\frac{\left[3+y^{2}+6\epsilon\sqrt{1+\chi^{2}}\left(1+y^{2}\right)^{2}\left(1+\chi^{2}\right)+2\chi^{2}\right]}{2\pi \left(1+\chi^{2}\right)^{3/2}\left(1+y^{2}\right)^{2} \left(\chi^{2}-y^{2}\right)^{-(1/2)}},
\end{equation}
where we have defined the dimensionless variables $y=R/a$ and $\tilde{\Sigma} = \frac{a^2\Sigma}{M}$. For the Hernquist model, we find that
\begin{eqnarray} \label{Hernsurf}
\tilde{\Sigma}(y)&=&\frac{1}{2\pi}\Bigg[\sqrt{\chi^2-y^2}\left[6\epsilon + \frac{-4-3\chi+y^2}{(1+\chi)^2 \left(1-y^2\right)^2}\right] \\ \nonumber
&&+\,\frac{2+y^2}{\left(1-y^2\right)^{5/2}} \ln\left(\frac{\left(1+\chi\right)y}{\chi+y^2-\sqrt{1-y^2}\sqrt{\chi^2-y^2}}\right)\Bigg].
\end{eqnarray}

If we let $\chi \rightarrow \infty$ and $\epsilon \rightarrow 0$, the familiar expression for the Plummer and Hernquist profiles \citep{ea05} are recovered. It is also possible to calculate the LOSVD for the Plummer and Hernquist models, but we shall not present the results of this calculation, as they are quite cumbersome. The cumulative surface brightness is yet another physically relevant expression that can be derived once $\Sigma$ has been determined. 

For the isotropic case with $\beta = 0$, one can evaluate the velocity dispersion profiles without explicitly computing the distribution function. We use the radial component of the spherical Jeans' equation, with $\beta=0$ and the boundary condition $\sigma_r^2=0$ when $x=\chi$, which on simplification reduces to
\begin{equation} \label{isoJeans}
\sigma_{r}^{2} = \frac{1}{\rho(r)} \int^\chi_r \rho\left(r'\right) \frac{d\Phi\left(r'\right)}{dr'}\,dr'.
\end{equation}

In general, this turns out to be a highly complicated expression for arbirtrary $p$.  For the Hernquist model, we obtain
\begin{eqnarray} \label{Hernisovel}
\frac{\sigma_{r}^{2}}{\Phi_{0}}&=&\frac{F_{H}\left(\chi\right)-F_{H}\left(x\right)}{x^{-1}\left(1+x\right)^{-3}+3\epsilon}, \\ \nonumber
F_{H}(x)&=&\Bigg[\frac{1}{4\left(1+x\right)^{4}}+\frac{1}{3\left(1+x\right)^{3}}+\frac{1-2\epsilon}{2\left(1+x\right)^{2}}+\frac{1-3\epsilon}{1+r} \\ \nonumber
&&-\,6\epsilon^{2}\left(1+x\right)+3\epsilon^{2}\left(1+x\right)^{2}+\ln\left(\frac{x}{1+x}\right)\Bigg].
\end{eqnarray}
Upon solving for the Plummer model, one has
\begin{eqnarray} \label{Plumisovel}
\frac{\sigma_{r}^{2}}{\Phi_{0}}&=&\frac{F_{P}\left(\chi\right)-F_{P}\left(x\right)}{\left(1+x^{2}\right)^{-(5/2)}+2\epsilon},
\end{eqnarray}
\[
F_{P}(x)=\frac{-1+12\epsilon^{2}\left(1+x^{2}\right)^{4}-4\epsilon\left(1+x^{2}\right)^{1/2}\left(4+7x^{2}+3x^{4}\right)}{6\left(1+x^{2}\right)^{3}}.
\]
Lastly, the case with $p=1/2$ yields
\begin{eqnarray} \label{NFWisovel}
\frac{\sigma_{r}^{2}}{\Phi_{0}}&=&\frac{F_{N}\left(\chi\right)-F_{N}\left(x\right)}{x^{-(3/2)}\left(1+\sqrt{x}\right)^{-4}+4\epsilon}, \\ \nonumber
F_{N}\left(x\right)&=&\Bigg[\frac{1}{3\left(1+\sqrt{x}\right)^{6}}+\frac{6}{5\left(1+\sqrt{x}\right)^{5}}+\frac{3}{\left(1+\sqrt{x}\right)^{4}} \\ \nonumber
&&-\,\frac{4\left(-5+\epsilon\right)}{3\left(1+\sqrt{x}\right)^{3}}+\frac{15-4\epsilon}{\left(1+\sqrt{x}\right)^{2}}+\frac{42}{\left(1+\sqrt{x}\right)}-\frac{1}{x} \\ \nonumber
&&+\,\frac{14}{\sqrt{x}}+4\epsilon^{2}x^{2}-56\ln\left(\frac{1+\sqrt{x}}{\sqrt{x}}\right)\Bigg].
\end{eqnarray}
By construction, it is evident that (\ref{Hernisovel}), (\ref{Plumisovel}) and (\ref{NFWisovel}) are zero when $x=\chi$. This is consistent with the assumption that our system is of finite extent, and does not extend beyond the cutoff radius.

Next, we consider the scenario where the distribution function, and consequently the velocity dispersions, are anisotropic. We make use of (\ref{rhoaug1}) in (\ref{sigmar2}) and (\ref{sigmat2}) to compute the velocity dispersion profiles. After some algebra, we find that
\begin{eqnarray} \label{sigmarmodel}
\frac{\sigma_r^2}{\Phi_0} &=& \Bigg[\frac{\frac{1}{2}x^{p-2}\left[\left(1+x^p\right)^{-(2p+2)/p}-(\epsilon x^2)^{2p+2}\right]}{(1+p)x^{p-2}\left(1+x^p\right)^{-2-(1/p)}+6\epsilon} \\ \nonumber
&&+\,\frac{\frac{6\epsilon}{1+\nu p}\left(1 + x^p\right)^\nu\left[\left(1+x^p\right)^{-(1+\nu p)/p}-(\epsilon x^2)^{1+\nu p}\right]}{(1+p)x^{p-2}\left(1+x^p\right)^{-2-(1/p)}+6\epsilon} \Bigg],
\end{eqnarray} 
\begin{eqnarray} \label{sigmatmodel}
\frac{\sigma_\theta^2}{\Phi_0} &=& \Bigg[\frac{\epsilon\left(1+p\right)x^{p}\left[\left(1+x^{p}\right)^{-(2p+1)/p}-\left(\epsilon x^{2}\right)^{2p+1}\right]}{(1+p)x^{p-2}\left(1+x^{p}\right)^{-2-(1/p)}+6\epsilon} \\ \nonumber
&&+\,\frac{\frac{p}{4}x^{p-2}\left[\left(1+x^{p}\right)^{-(2p+2)/p}-\left(\epsilon x^{2}\right)^{2p+2}\right]}{(1+p)x^{p-2}\left(1+x^{p}\right)^{-2-(1/p)}+6\epsilon} \\ \nonumber
&&+\,\frac{\frac{3\epsilon\nu p}{\nu p+1}x^{p}\left(1+x^{p}\right)^{\nu-1}\left[\left(1+x^{p}\right)^{-(\nu p+1)/p}-\left(\epsilon x^{2}\right)^{\nu p+1}\right]}{(1+p)x^{p-2}\left(1+x^{p}\right)^{-2-(1/p)}+6\epsilon} \\ \nonumber
&&+\,\frac{\frac{6\epsilon}{\nu p+1}\left(1+x^{p}\right)^{\nu}\left[\left(1+x^{p}\right)^{-(\nu p+1)/p}-\left(\epsilon x^{2}\right)^{\nu p+1}\right]}{(1+p)x^{p-2}\left(1+x^{p}\right)^{-2-(1/p)}+6\epsilon} \\ \nonumber
&&+\,\frac{6\epsilon^{2}x^{2}\left(1+x^{p}\right)^{\nu}\left[\left(1+x^{p}\right)^{-(\nu p)/p}-\left(\epsilon x^{2}\right)^{\nu p}\right]}{(1+p)x^{p-2}\left(1+x^{p}\right)^{-2-(1/p)}+6\epsilon} \Bigg].
\end{eqnarray}
From the above expressions, it is clear that the velocity dispersions vanish when $x=\chi$, akin to the isotropic case, when one uses the definition of $\chi$. In addition, the positivity of all the variables ensures that the resultant velocity dispersions are also positive. The anisotropy parameter can be evaluated in a similar manner by using (\ref{betafromv}).

	\label{lastpage}

\end{document}